\documentclass[submission,copyright,creativecommons]{eptcs}
\providecommand{\event}{SOS 2007} 

\usepackage{iftex}

\ifpdf
  \usepackage[T1]{fontenc}        
\else
  \usepackage{breakurl}           
\fi

\title{A Prolog Program for Bottom-up Evaluation}
\author{David S. Warren
  \institute{Department of Computer Science\\
    Stony Brook University\\ Stony Book, New York}
\email{warren@cs.stonybrook.edu}}

\def\titlerunning{Prolog for Bottom-up}
\def\authorrunning{D.S. Warren}
\begin{document}
\maketitle

\begin{abstract}
This short paper describes a simple and intuitive Prolog program, a metainterpreter, that computes the bottom up meaning of a simple positive Horn clause definition.  It involves a simple transformation of the object program rules into metarules, which are then used by a metainterpreter to compute bottom up the model of the original program.  The resulting algorithm is a form of semi-naive bottom-up evaluation.  We discuss various reasons why this Prolog program is particularly interesting.  

    In particular, this is perhaps the only Prolog program for which I find the use of Prolog's {\tt assert/1} to be intrinsic, easily understood, and the best, most perspicuous, way to program an algorithm.  This short paper might be best characterized as a Prolog programming pearl.
\end{abstract}

\section{Introduction}

A positive Prolog program is an inductive definition of a set of relations over terms.  There are two well-known basic ways to compute elements of relations that are defined in this way \cite{DBLP:journals/jacm/EmdenK76}, \cite{DBLP:books/sp/Lloyd84}: top-down \cite{DBLP:conf/ifip/Kowalski74} and bottom-up \cite{DBLP:conf/db-workshops/Bancilhon85}.

Top-down evaluation starts with a query to a defined relation that asks for members of the defined relation that are instances of the query.  It is query driven, since at each step a set of subqueries is posed that drive the search for an answer.  Bottom-up evaluation starts with the facts, the data, and uses existing facts to iteratively infer new facts to ultimately compute all members of the defined relations.  

Prolog uses SLD resolution, a top-down, query-driven evaluation strategy using highly optimized compilation techniques.  Datalog systems use a bottom-up, data-driven evaluation strategy with engines optimized for that execution.  Some problems are more efficiently solved top down and some bottom up.  Thus there are some problems for which a Prolog programmer would like to use the Prolog built-in top-down strategy to perform bottom-up evaluation.

Several Prolog systems now support a hybrid evaluation strategy called {\em tabled evaluation} (e.g., \cite{XSB},\cite{Yap},\cite{SWI},\cite{Ciao}), which allows those Prolog systems to use aspects of bottom-up evaluation within the Prolog top-down framework.  Tabling normally can provide all the bottom-up processing needed by a Prolog programmer.  But some Prolog systems do not have tabling, and tabling is not precisely bottom-up evaluation in any case.

Another use for bottom-up processing in Prolog is for teaching and debugging purposes.  Students can learn more about what their Prolog definitions mean by exploring how they would evaluate bottom up.  And experienced programmers may be able to find why Prolog does not compute an expected result by exploring the bottom-up computation.

For all these reasons a good, simple, customizable implementation of bottom-up evaluation within a Prolog environment is desirable.  Prolog is well-known for the elegance of the metainterpreters that it supports, and a bottom-up evaluator requires a metainterpreter.  But Prolog metainterpreters normally inherit a top-down strategy from the Prolog evaluator.  A metainterpreter that interprets object programs bottom up is not so obvious.

This short paper describes a metainterpreter written in (standard ISO) Prolog that evaluates pure, positive Prolog programs bottom up.

\section{The Algorithm}
The first step of the algorithm is to transform the object program into a new set of rules, rules that define a new binary predicate we call {\tt implies/2}.  The {\tt implies} rules are generated from the object program as follows:

For each object rule: \[H \mbox{ :- } B_1, B_2, ..., B_n.\] generate $n$ {\tt implies/2} rules, one for each $B_i$.  Specifically, for each $B_i$ generate  \[\mbox{implies}(B_i,H) \mbox{ :- } B_1, B_2, ..., B_{i-1}, B_{i+1}, ... B_n.\]
And for each object fact of the form $H.$ generate an {\tt implies/2}  fact of the form $\mbox{implies}(\mbox{true},H).$

The metainterpreter uses these generated {\tt implies/2} rules to evaluate the original object program bottom up.  The main metainterpreter predicate {\tt bu(Schedule,SchedEnd)} takes a difference list that contains all the facts that have been derived but have not yet been used, along with the rules and other known facts, to derive new facts.

The Prolog code for the metainterpreter {\tt bu/2} is:
\begin{verbatim}
bu([],[]).          % finished
bu([Fact|Sched],SchedEnd) :-
    nonvar(Fact),  % another derived fact to process
    findall(Hd,implies(Fact,Hd),Inferred),  % get newly inferred facts
    assert_and_sched_inferred(Inferred,SchedEnd,NewSchedEnd),
    bu(Sched,NewSchedEnd).
\end{verbatim}
The first clause terminates when there are no more derived facts whose inferences are to be propagated.  The second clause is the workhorse: it takes the next fact from the schedule, calls {\tt implies/2} on that fact and collects together the heads of all object rules now satisfied using that fact, i.e. facts that can now be inferred.  It then uses {\tt assert\_and\_sched\_inferred} to process all those facts: for each that is not already in the database, it adds it to the database and adds it to the end of the schedule.  And iterates.

The supporting predicates are:
\begin{verbatim}
assert_and_sched_inferred([],Sched,Sched).
assert_and_sched_inferred([Fact|Inferred],Sched,SchedEnd) :-
    (subsuming_fact(Fact)
     -> assert_and_sched_inferred(Inferred,Sched,SchedEnd)
     ;  assert(Fact),
        Sched = [Fact|SchedTail],
        assert_and_sched_inferred(Inferred,SchedTail,SchedEnd)
    ).
\end{verbatim}
\begin{verbatim}
subsuming_fact(Hd) :-
    term_variables(Hd,Variables),
    call(Hd),
    is_most_general_term(Variables).  % all still variables and distinct
\end{verbatim}
The initial call to generate the least fixpoint is:
\begin{verbatim}
?- bu([true|End],End).
\end{verbatim}

Consider the meaning of a generated rule for {\tt implies/2} in {\tt bu/2}. Atom {\tt implies(A,H)} is true if there is a rule in the object program that has head {\tt H} and a body literal {\tt A} and all the other body literals of the rule are true in the database.  I.e., {\tt A} implies {\tt H} if the remaining body literals are true.  This is clearly a consequence of the original object program rule.  These rules are applied starting with the empty database, finding facts implied by the database facts and these rules, updating the database with the newly inferred facts, and iterating until no new facts can be inferred.

\section{Transitive Closure Example}

Consider the Prolog rules for transitive closure of a simple graph:
\begin{verbatim}
    tc(X,Y) :- edge(X,Y).
    tc(X,Y) :- edge(X,Z), tc(Z,Y).

    edge(a,b).
    edge(b,c).
    edge(c,b).
\end{verbatim}
We note that since the graph in {\tt edge/2} is a cyclic graph, Prolog will go into an infinite loop when asked the query \verb|:- tc(A,B).|  But bottom-up evaluation will terminate as we shall see.

The {\tt implies/2} rules generated from these object rules and facts are:
\begin{verbatim}
    implies(edge(X,Y),tc(X,Y)).

    implies(edge(X,Z),tc(X,Y)) :- tc(Z,Y))
    implies(tc(Z,Y),tc(X,Y)) :- edge(X,Z))

    implies(true,edge(a,b)).
    implies(true,edge(b,c)).
    implies(true,edge(c,b)).
\end{verbatim}
Each iteration of the main loop takes a fact from the schedule and produces a list of newly derived facts.  that trace is:
\begin{verbatim}
  true adds [edge(a,b),edge(b,c),edge(c,b)]
  edge(a,b) adds [tc(a,b)]
  edge(b,c) adds [tc(b,c)]
  edge(c,b) adds [tc(c,b),tc(c,c)]
  tc(a,b) adds []
  tc(b,c) adds [tc(a,c)]
  tc(c,b) adds [tc(b,b)]
  tc(c,c), tc(a,c), and tc(b,b) each adds []
  finished
\end{verbatim}

This evaluator also handles unsafe programs, i.e., those with variables in answers.  For example:
\begin{verbatim}
    append([],L,L).
    append([X|L1],L2,[X|L3]) :- append(L1,L2,L3).
\end{verbatim}
which generates the {\tt implies/2} facts:
\begin{verbatim}
    implies(true,append([],A,A).
    implies(append(A,B,C),append([D|A],B,[D|C]).
\end{verbatim}
and its trace:
\begin{verbatim}
  true adds [append([],A,A)]
  append([],A,A) adds [append([B],C,[B|C])]
  append([B],C,[B|C]) adds [append([D,E],F,[D,E|F])]
  append([D,E],F,[D,E|F]) adds [append([G,H,I],J,[G,H,I|J])]
  and it continues...
\end{verbatim}
Clearly if a Prolog program has infinitely many answers, its bottom-up evaluation will not terminate.  But it can be useful to see the first several answers, as we've done here for {\tt append/3}.  We can modify our bottom-up evaluator to return nondeterministically each fact and the new facts it generates.  This allows the user to see the facts as they are iteratively generated and either stop the generation or to ask for another.  The following modification of {\tt bu/2} is such a program:
\begin{verbatim}
  bu([Fact|Sched],SchedEnd,Fact1,NewFacts) :-
      nonvar(Fact),  % done if a variable.
      findall(Hd,implies(Fact,Hd),Inferred),
      assert_and_sched_new(Inferred,SchedEnd,NewSchedEnd),
      (NewSchedEnd = [], SchedEnd \== [], % something added
       Fact1 = Fact, New = SchedEnd  % return this generator and generated set
       ;
       bu(Sched,NewSchedEnd,Fact1,NewFacts) % on to more, if asked
      ).
\end{verbatim}

\section{Why this Prolog Program Is Interesting}

Why might this problem of bottom-up evaluation make for an interesting Prolog program?
\begin{enumerate}
\item It solves a basic problem in logic.  This implementation in Prolog may provide evidence for a sometimes made claim that the Prolog programming language is a good assembler language, or implementation language, for logic-based problems.
\item It is a fundamental problem in the theory of logic programming.
\item It is a metainterpretation problem, for which Prolog is known for elegantly solving.
\end{enumerate}

Prolog programs using {\tt assert/1}, as this one does, do not have a declarative meaning and in general can be notoriously difficult to understand.  This Prolog program is fairly easy to understand even though it uses {\tt assert/1} in a fundamental way and does not therefore have a traditional declarative meaning.  Nevertheless, it is perspicuous for several reasons:
\begin{enumerate}
\item The program is essentially deterministic.  The persistence of asserted clauses over backtracking is one feature of {\tt assert/1} that can make programs using it obscure.  This program has essentially no backtracking.
\item The primary structure of this program is a single deterministic loop, clearly indicated by its tail recursive call.  Each time around the loop adds a new set of facts to the database.
\item The nondeterminism, which is very convenient, is encapsulated by the {\tt findall/3}) metacall.  It simply collects the set of facts to be added to the database on the current iteration, which are nondeterministically generated by the call to the {\tt implies/2} predicate.
\item The use of {\tt assert/1} allows us to use Prolog's rule evaluation mechanism to determine what new atoms can be inferred on an iteration.  The call to Prolog predicate {\tt implies/1}, implicit in the {\tt findall/3}, performs a metainterpretation step, at the object level.
\item The logical meaning of the {\tt implies} clauses is clear and compelling.
\end{enumerate}

These observations lead me to believe that this is an interesting Prolog program, even though it inherits no declarative meaning from Prolog's logical semantics.  Its meaning comes from a combination of Prolog's procedural meaning (for the deterministic loop that updates the database) and its declarative meaning (for the call to {\tt implies/2} and its accumulation with {\tt findall/3}).

\section{Complexity}
On its face, this program seems efficient.  It doesn't recompute in each step everything that was computed in the previous step as naive bottom-up evaluations does.  I.e., it performs what is known as a semi-naive computation.  Every call to {\tt implies/2}, the workhorse of the computation, will use the standard first-argument indexing of all Prolog evaluators (but it may not be optimally indexed).  And those calls use Prolog's built-in evaluation mechanism and so can avoid a level of metainterpretation.  All this bodes well for good performance. 

However, we can see a possible source of redundant computation.  Consider an object program clause that has a large number of body atoms: $H \mbox{ :- } B_1, B_2, ..., B_n$ for large $n$.  There will be $n$ {\tt implies/2} clauses generated from it.  Assume that the rule does indeed fire at some iteration. That will be after every $B_i$ has been asserted into the database at some iteration.  Say they are added in their left-to-right order, i.e., $B_1$ is added first, then $B_2$ at a later iteration, then $B_3$ at a sill later iteration, etc., until finally $B_n$ is added and the {\tt implies/2} rule for $B_n$ causes the head to be added to the database.  Notice that the first body atom $B_1$ will have been called $n-1$ times, once at each level that some $B_i$ was added.  And $B_2$ will be called $n-2$ times, once for each level after the first has been added.  And so on.  So there is $O(n^2)$ computation when only $O(n)$ is theoretically needed.  (One could imagine removing a body literal when it becomes true.  How to do this efficiently with the data structures used here is not so obvious.)  Indeed the $n$ is only the maximum number of literals in the body of any clause, which for normal programs is not very big.  But it is still an unnecessary redundancy and for some programs may be very costly.

This happens only for programs with clauses with many body literals.  If all program clauses have at most two body literals, then this redundancy does not occur.  And any program can be transformed into one in which all rules have two or fewer body literals by a folding process that introduces intermediary predicates.  E.g., $H\mbox{ :- }B_1, B_2, B_3$ can be replaced by two rules $H\mbox{ :- }H', B_3$ and $H'\mbox{ :- }B_1, B_2$ where $H'$ is new.  This process reduces the number of literala of a rule by one, and can be iterated until we have at most two body literals for any rule.  This program transformation was introduced in early Datalog systems, which are bottom up, for exactly this reason.  

We noted that the {\tt implies/2} predicate is by default indexed on its first argument, which is bound for every call.  However, if the index is on only the main functor symbol, then the effective indexing is only on the predicate name of the newly added fact.  Some Prolog's support deeper indexing, which would be helpful here.

\section{Generating the {\tt\bf implies/2} Rules}

In this section we provide definitions for a couple of supporting functions: the generation of {\tt implies/2} clauses from Prolog clauses, and the conversion of clauses with more than two body literals into a set of clauses with at most two by the introduction of intermediary predicates and folding.

The first predicate takes a list of Horn clauses and asserts the corresponding set of {\tt implies/2} clauses.  Other variations of this functionality might be more appropriate for integration into a larger system.  For example, using a ``term expansion'' macro functionality of some Prologs might be effective.

\begin{verbatim}
    assert_imply_rules([]).
    assert_imply_rules([(Head:-Body)|Rules]) :-
        (do_all
         delete_one(Atom,Body,NBody),
         assert((implies(Atom,Head) :- NBody))
        ),
        assert_imply_rules(Rules).

    delete_one(G1,G1,true) :- \+ G1 = (_,_).
    delete_one(G1,(G1,G2),G2).
    delete_one(G2,(G1,G2),G1) :- \+ G2 = (_,_).
    delete_one(SG,(G1,G2),(G1,G3)) :-
        G2 = (_,_),
        delete_one(SG,G2,G3).
\end{verbatim}
The predicate {\tt delete\_one/3} is like Prolog's {\tt select/3} (sometimes called {\tt delete/3}) for lists but works on comma-structures.  The messy tests are because of their ``default-y'' representation \cite{10.5555/83204} and to avoid an extraneous {\tt true} goal at the end a clause.

The second functionality is the generation of folded clauses to provide better complexity for our bottom-up metainterpreter.  For most applications envisioned this optimization may be unnecessary and in fact may complicate the understanding intended to be enhanced by exploring bottom-up evaluation.  We also note that the choice of lierals to fold out and to keep and their order may affect the efficiency of the resulting rules.  We give this simple definition here just for completeness. 

The predicate {\tt expand\_rules} takes a list of clauses and returns an equivalent list of clauses each with two or fewer body literals:
\begin{verbatim}
    expand_rules([],[]).
    expand_rules([Rule|Rules],[NRule|ExpRules]) :-
        (gen_folded(Rule,NRule,TRule)
        -> expand_rules([TRule|Rules],ExpRules)
        ;  NRule = Rule,
           expand_rules(Rules,ExpRules)
        ).


    gen_folded((Head :- Lit1,TBody),(Head:-Lit1,THead),(THead:-TBody)) :-
        TBody = (_,_),
        gensym('_$Tmp',Pred),
        Outer = (Lit1,Head),
        excess_vars(Outer,TBody,[],Vs1),
        excess_vars(TBody,Outer,Vs1,Vs2),
        excess_vars((Outer,TBody),Vs1-Vs2,[],Vs),
        THead =.. [Pred|Vs].
\end{verbatim}

\section{Conclusion}

We have presented a metainterpreter written in Prolog to evaluate Horn clause programs bottom up.  We find this program interesting for its simplicity and clarity.  To understand it, one must understand part as a procedural computation and part as a declarative specification and integrate those understandings.  To my mind it is unusual in that Prolog's {\tt assert/1} operation is intrinsic and adds to the program's elegance.


\bibliographystyle{eptcs}
\bibliography{buinprolog}
\end{document}